# SOCIO-ECONOMICAL ANALYSIS OF ITALY: THE CASE OF HAGIOTOPONYM CITIES

Roy Cerqueti
University of Macerata, Department of Economics and Law
Via Crescimbeni 20, I-62100, Macerata, Italy
Tel.: +39 0733 258 3246; Fax: +39 0733 258 3205
E-mail: roy.cerqueti@unimc.it

Marcel Ausloos
School of Management, University of Leicester
University Road, Leicester, LE1 7RH, UK
E-mail: ma683@le.ac.uk

eHumanities Group
Royal Netherlands Academy of Arts and Sciences
Joan Muyskenweg 25, 1096 CJ Amsterdam, The Netherlands

GRAPES – Group of Researchers for Applications of Physics in Economy and Sociology
Rue de la Belle Jardiniere 483, B-4031, Angleur, Belgium
E-mail: marcel.ausloos@ulg.ac.be

**Abstract**

This paper pursues the scopes of joining the economical characteristics of Italian cities with a relevant sociological aspect: the cult of the catholic Saints. Indeed, more than in other Countries, a high percentage of Italian cities has a toponym coming from the name of specific Saints (hagiotoponym). The assessment of the historical origin of each hagiotoponym is out of the scopes of the present paper, but the link with the religious sense of Italians seems to be clear. The statistical analysis of the economic contributions that each hagiotoponym city provides to the Italian GDP is here performed. Such an analysis is also based on the comparison with the overall Italian data, and it is carried out through the computation of the Theil, Gini and Herfindahl-Hirschman indices.

1. Introduction

In New Economic Geography, it is attempted to examine the relationships between specifically defined entities and the global and local economies. The wealth flow wealth is one item often considered, beside the various political and geographic constraints.

In so doing, the problem of assessing the statistical properties of geographic-economic structure of a country has been widely debated in the last decades (Fan and Scott, 2003, and references therein).

However, economic indicators, other than histograms and/or rank-size plots, can be useful. The Theil, Herfindahl-Hirschman and Gini indices as measures of various economic specificities are considered below. Here, their calculation are adapted to cities in a country. The concerned distribution relies on the aggregated tax income (ATI) of a city, and the considered period is 2007-2011. The statistical indicators here employed are now listed.

First of all, consider $N$ cities and denote as $y_i$ the ATI of the $i$-th city, for each $i=1,2,...,N$, so that the ratio e 5 yearly official data for the [2007-2011] time interval.

Next, it should be realized that the cluster selection needs much research and many considerations of different types. The data collection relies on the identification of Italian cities having a toponym recalling a Saint. Such an identification is a tedious stage, and is implemented in several phases. Firstly, a preliminary list of cities is constructed through the application of four sorting : (i) by employing the string "sa", (ii) by extracting from the previous list the city names containing "San ", "Santo ", "Santa ", "Sant' ',' with the blank space at then end of the strings, (iii) by removing the string "san", without space after the word, from the previous list, and (iv) by examining all such cases carefully in order to be sure that the hagiotoponym was a human Saint. For example, we reject cities like Camposanto, Sant'Angelo, (24 times), Santa Croce (5 times), Santa Luce, Santo Spirito, and similar non-human cases. Cities containing Notre-Dame and Madonna are also added to the list. Moreover, after much thought, Michele (11 times) and Raffaele (once) are added, although they are not humans, but archangels. However, they are so much anthropomorphic that they can be here assimilated to human Saints. This result in a list of 637 municipalities.

## 2. Results

The usual statistical characteristics of the distribution of ATI values, of the 8092 Italian cities and of the 637 Italian hagiotoponyms are reported in Table 1.

The values of the economic indices are given in Table 2. At the country level, the Gini coefficient is rather high: it points to a high level of disparity among Italian cities. When considering the hagiotoponyms cluster, Gini index becomes remarkably smaller. This means that ATI is fairly distributed among hagiotoponyms. The reason for this discrepancy should be found in the presence of outliers only in the whole Italian cities distribution.

The *HHI* index for hagiotoponyms also indicates the lack of disparities. Not only, the index is low for the whole country, indicating, a small difference in ATI value of cities toward the Italian GDP, but also that the fact that a Saint name in the name of a city can be associated to a not too much scattered values. For such a catholic country, this is somewhat expected, being the cult of the Saints widespread across the entire territory. Therefore, the Saint cities reflect the national reality of overall medium-sized urban agglomerations without having outliers like Roma, Milano and similar huge urban areas. This finding is also confirmed by the Theil index, which is much smaller for the hagiotoponym samples than for the entire Italy. This indicates a greater homogeneity in the *Hagiotop.* set. However, the fact that it occurs for the three indices is not *a-priori* expected. Recall that the *HHI* index, the *Th* and *Gi* indices allow to gain different insights onto the distributions of measures (ATI). On one hand, *HHI* based on the second moment of the measurement distributions informs on the distribution variance, thus the spread of the level of competition among cities, whence on their interactions, while both *Th* and *Gi* indices provide measures of the data dispersion. In fact, *Th* baed on an entropy idea, i.e. the zero moment of the probability distribution, leads to a number which synthesizes the degree of probable dispersion of an agent in the population.

| Statistical indicator of $<ATI>_{5yr}$ | (i) | (ii) |
|---|---|---|
| | *Italian cities* | *Hagiotop.* |
| Minimum | 332190 | 775239 |
| Maximum | 4.47e+10 | 1.24e+09 |
| Points | 8092 | 637 |
| Mean | 87391347 | 54719782 |
| Median | 23827615 | 23627546 |
| RMS | 6.68e+08 | 1.10e+08 |
| Std Deviation | 6.62e+08 | 96604556 |
| Variance | 4.39e+17 | 9.33e+15 |
| Std Error | 7364451.9 | 3827611.9 |
| Skewness | 49.14 | 5.69 |
| Kurtosis | 2956.78 | 47.75 |

Table 1. Statistical characteristics of the distribution of ATI values, of (i) the 8092 Italian cities; (ii) the 637 hagiotoponym Italian cities.

3. **Conclusions**

This paper offers a socio-statistical analysis of the Italian economic situation. In particular, due to the catholic nature of the considered country, the statistical properties of the ATI of the Italian cities which contain -in the toponym- a reference to a human Saint are discussed. It seemed that relevant scientific questions could be raised on the pertinence of socio-economic features due to a saint name defining a city. Results are quite interesting: the hagiotoponym cities can be viewed as a proxy of the overall national reality of medium-sized cities, when outliers are removed.

| Economic indicator of $<ATI>_{5yr}$ | (i) | (ii) |
|---|---|---|
| | *Italian cities* | *Hagiotop.* |
| $N$ | 8092 | 637 |
| $\ln(N)$ | 8.9986 | 6.4568 |
| $H$ | 7.2650 | 5.6914 |
| $Th$ | 1.734 | 3.144 |
| $HHI \times 10^2$ | 0.7332 | 0.6455 |
| $H^* \times 10^2$ | 0.7209 | 0.4893 |
| $Gi$ | 0.7591 | 0.62407 |

Table 2. Economic indicator of the distribution of ATI values, of (i) the 8092 Italian cities; (ii) the 637 hagiotoponym Italian cities. It is worth noting that the maximum level of entropy is ln (*N*).